\documentclass[journal,twocolumn]{IEEEtran}

\usepackage{subfigure}
\usepackage[dvips]{graphicx}
\usepackage{color}
\usepackage{soul}
\usepackage{cite}
\usepackage{lineno}
\usepackage{multirow}
\usepackage{upgreek}
\usepackage{tabulary}
\usepackage{multicol}
\usepackage{mathtools}
\usepackage[T1]{fontenc}
\usepackage{amsmath}

\usepackage{lipsum}

\begin{document}

\title{Classification of Breast Lesions Using Quantitative Ultrasound Biomarkers}

\author{~Navid Ibtehaj Nizam, Sharmin R. Ara, and Md. Kamrul Hasan
\thanks{N. I. Nizam,  and M. K. Hasan (Corresponding Author, khasan@eee.buet.ac.bd) are with the Department
of Electrical and Electronic Engineering, Bangladesh University of Engineering and
Technology, Dhaka-1205, Bangladesh.}
\thanks{S. R. Ara is with the Department
of Electrical and Electronic Engineering, East West University, Dhaka-1212, Bangladesh.}}

\maketitle

\begin{abstract}
Quantitative ultrasound (QUS) based parameters like the effective scatterer diameter (ESD) and mean scatterer spacing (MSS) are gaining attention recently as non-invasive biomarkers for soft tissue characterization. In this work, we propose a multiple QUS parameter based technique that employs ESD and MSS, for binary classification of breast lesions. In order to produce improved ESD estimates, we propose a modified frequency domain technique for ESD estimation of breast tissues from the diffuse component of backscattered radio-frequency (RF) data. Ensemble empirical mode decomposition (EEMD) is performed to separate the diffuse component from the coherent component by decomposing the RF data into their intrinsic mode functions (IMFs). A non-parametric Kolmogorov-Smirnov (K-S) test is employed for automatic IMF selection along with a multi-step system effect minimization process. The ESD is estimated using a nearest neighborhood average regression line fitting algorithm. Furthermore, we use an ameliorated EEMD domain autoregressive (AR) spectral estimation technique for MSS estimation. On using the ESD for binary classification of $159$ lesions, we obtain high sensitivity, specificity, accuracy values of $91.07\%$, $96.12\%$, and $94.34\%$, respectively, with an area under the receiver operating characteristics (ROC) curve of $0.94$. On combining ESD with MSS we obtain even more improved sensitivity, specificity, and accuracy values of $96.43\%$, $95.15\%$, and $95.60\%$, respectively, with an area under the ROC of $0.96$. Such a high classification performance highlights the potential of these QUS parameters to be used as non-invasive biomarkers for breast cancer detection.
\end{abstract}

\IEEEpeerreviewmaketitle



\section{Introduction}
\label{intro}

Quantitative ultrasound (QUS) based parameters have been employed for the characterization and classification of soft tissues with a view to developing a diagnostic modality which is less subjective to operator settings and interpreter variability compared to diagnosis based on conventional ultrasound imaging \cite{Nam_2013}, \cite{Bige_05}, \cite{Tadayyon_14}. Moreover, QUS parameter based breast lesion classification provides some inherent advantages compared to diagnosis based on other modalities, which are commonly employed for breast cancer detection. These advantages result from the non-ionizing and non-invasive nature of ultrasound (advantageous over mammography and needle biopsy) and lower cost of operation (advantageous over magnetic resonance imaging) \cite{Alacam_04}. The main idea behind the use of QUS for tissue characterization is that the disease processes alter the physical properties of tissues and this, in turn, is reflected by a change in the acoustic scattering properties of tissues \cite{Lizzi_86}.

The potential of effective scatterer diameter (ESD) and mean scatterer spacing (MSS) as QUS parameters for classification and characterization of pathological tissues have been reported extensively in the literature. These parameters, may be referred to as microscopic QUS parameters, since they are not observable from ultrasound B-mode or ultrasound elastography images but, rather, need to be estimated from histopathology slides using microscopy. The ESD of ocular tumors \cite{Felleppa_86},  liver \cite{Liu_2007}, renal tissues \cite{Insana_93_IR}, glomerular tissues \cite{Hall_96}, kidney \cite{Insana_95}, prostate \cite{Feleppa_97}, human aortae \cite{Bridal_97}, and the uveal melanomas \cite{Silverman_97} have been reported.  Additionally, ESD has been successfully used to distinguish between benign and malignant lesions in the eyes \cite{Liu_2004} and in the lymph nodes \cite{Mamou_2010}. Ultrasonic characterization of human breast tissues, based on ESD, has been reported in \cite{Tadayyon_14}. However, \cite{Tadayyon_14} used ESD along with mean scatterer spacing (MSS) and effective acoustic concentration (EAC) for tumor grading rather than breast lesion classification. Our group previously used mean scatterer spacing (MSS) for classification of breast lesions on a small dataset using an ensemble empirical mode decomposition (EEMD) domain autoregressive (AR) spectral estimation technique \cite{Nizam_17}. A very recent work in \cite{Nasief_19} used multiple QUS parameters for classification of breast masses but that too, was done on a limited dataset. To the best of our knowledge, no previous work has been reported on a large dataset, for classifying between benign and malignant breast lesions using microscopic QUS biomarkers like ESD and MSS, either individually or combined.

QUS methods exploit the frequency dependence of backscattered radio frequency (RF) signals for the estimation of attenuation of sound waves in tissues and model the backscattered data to estimate  scatterer parameters such as the size, shape, spacing and number density in tissues \cite{Oleze_02}. Previously, it has been established that the backscattered RF data consists of a coherent component and a diffuse component \cite{Wear_93}. In \cite{Lavarallo_12}, it has been reported that, for the estimation of ESD, the coherent component behaves as an interference and hence, needs to be suppressed. The generalized spectrum and Rayleigh envelope statistics \cite{Luchies_12} and Hanning tapers \cite{Luchies_15} have been used to separate the diffuse echoes from the backscattered  data but these algorithms remain untested on human tissue. Recently, EEMD has been successfully employed to separate the coherent and diffuse components from deconvolved backscattered RF data for MSS estimation of female breast tissues \cite{Nizam_17}. ESD estimation techniques proposed previously often employ a Gaussian form factor to model tissue scattering \cite{Oleze_02}, \cite{Zachary_02}. A technique that relies on the minimization of the average squared deviation (MASD) between the theoretical and measured power spectra using a Gaussian form factor model has been proposed in \cite{Oleze_02}. However, this method does not employ any signal decomposition technique for separation of the coherent component from the diffuse component. A frequency domain technique that employs a Gaussian form factor based theoretical power spectrum has been proposed in \cite{Zachary_02}. This method also does not use any signal decomposition technique. Furthermore, the existing techniques use large $2$-D spatial signal blocks to generate a stable block power spectra \cite{Oleze_02}, \cite{Zachary_02}, which is often an unrealistic approach because the tissue pathology inside a large spatial region must be considered uniform, a requirement that is too idealistic for a heterogeneous tissue medium, and also for smaller-sized lesions.  To combat this problem, an attenuation estimation technique has been developed in \cite{Hasan_2013} which uses smaller spatial areas in the lateral direction.
\newline

The contribution of this work is two-fold:

\begin{enumerate}

    \item We propose an ESD and MSS based binary (benign-malignant) classification technique for $159$ breast lesions using linear classifiers. The individual classification performances of ESD and MSS are also considered.
    \item  We propose a new ESD estimation technique, based on a frequency domain Gaussian form factor model \cite{Zachary_02}, for improved classification of breast lesions using EEMD and a non-parametric Kolmogorov-Smirnov (K-S) \cite{Georgiou_01} test to automatically select the IMFs that show diffuse scattering. Furthermore, we adopt a multi-step system effect minimization process and a nearest neighborhood average regression line fitting (NNARLF) algorithm for accurate ESD estimation.   We also modify an existing EEMD domain autoreggresive (AR) spectral estimation based MSS estimation technique \cite{Nizam_17} so that it can be combined with ESD for binary classification on the same dataset. The motivation behind combining ESD with MSS is that the MSS has previously been shown to be successful in binary classification of breast lesions \cite{Nizam_17}. Additionally, the MSS can be estimated using a scheme similar to the proposed ESD estimation technique and also, the use of multiple QUS parameters is a common approach for ultrasonic tissue characterization  \cite{Tadayyon_14}, \cite{Nasief_19}, \cite{Golub_93}, \cite{Ara_2017}. A greater focus is placed on the technique for ESD estimation because there are significant modifications from the original frequency domain technique, \cite{Zachary_02}, on which our proposed ESD estimation technique is based. Therefore, numerical results are produced to highlight the important features of the proposed ESD estimation technique. Also, comparisons with some of the existing techniques are provided, for both ESD and MSS \cite{Tadayyon_14}, \cite{Oleze_02}, \cite{Zachary_02}, \cite{Pereira_01}.
\end{enumerate}

This paper is structured as follows. Section $2$ explains the proposed methods for ESD and MSS estimation. Section $3$ presents the ESD and MSS estimation results along with the classification results obtained from our proposed QUS estimators and compares the results with some existing techniques. Section $4$ discusses the important aspects of the proposed method and focuses on its strengths and limitations. Finally, Section $5$ sums up the paper with some concluding remarks.

\section{Materials and Method}
\subsection{Patient Data}
The \textit{in vivo} data used in this paper have been obtained at the Bangladesh University of Engineering and Technology (BUET) Medical Center with the help of a SonixTOUCH Research (Ultrasonix Medical Corporation, Richmond BC, Canada) scanner integrated with a L14--5/38 linear probe. The probe was operating at $10$ MHz with $65$\% bandwidth ($-6$-dB bandwidth) at a sampling rate of 40 MHz. The pulse length was approximately $0.4$ mm and the beam width was approximately $2.2$ mm. The pulse length has been estimated from the emitted pulse using the multiple input-output inverse theorem \cite{Miyoshi_88}. The study has been carried out on $179$ female subjects, with their prior written constant, and the study was approved by the Institutional Review Board (IRB) of BUET. ESD estimation is carried out on $245$ RF data records. Out of the $245$ data records, $56$ are malignant, $79$ are fibroadenomas, $24$ are inflammatory lesions, $42$ are cystic lesions, and $44$ are normal (that is, no lesions are present). However, binary classification of breast lesions is carried out on $159$ RF records excluding the cystic lesions and normal tissues. The cysts are excluded since ESD and MSS cannot be reliably estimated from them, as elaborated in the experimental results and discussion section. The age range of all patients was 13--75 years (mean: 35.27 years). The patients having masses underwent fine-needle aspiration cytology (FNAC) and/or excision biopsy according to the suggestion of their physicians. All patients having FNAC diagnosis positive for malignancy underwent surgery. Therefore, diagnoses of malignant and some benign lesions were confirmed by histopathology, and diagnoses of the remaining lesions, by cytopathology. The details of the RF data recorded from the breast tissues are summarized in Table \ref{patient_summary}.

\label{Methods}
\begin{table}[!ht]
\centering
\caption{Patient data summary}
\scriptsize
\resizebox{\columnwidth}{!}{
\begin{tabular}{ccccc}
\\
\hline   Tissue type  &   No.of &  Mean age&    Method of &   Lesion size\\
 & \ RF data  &     $\pm$SD &   confirmation &  mean(mm$\times$mm) $\pm$SD(mm$\times$mm)\\
\hline   Malignant &  56 &   44.91$\pm$9.89 &   Biopsy &   20.82$\times$19.06$\pm$5.78$\times$6.89\\
  Fibroadenoma &  79 &  27.48$\pm$9.13 &   FNAC/Biopsy  &   19.02$\times$11.45$\pm$7.93$\times$4.79\\

  Cyst     &  42  &  39.13$\pm$8.90 &   FNAC  &   13.12$\times$8.91$\pm$6.91$\times$4.83\\

  Inflammation  &  24  &  35.43$\pm$12.71 &   FNAC/Biopsy  &   21.53$\times$13.15$\pm$8.39$\times$4.39\\
   Normal  &  44  &  - &  Not applicable  &  Not applicable\\
\hline
\label{patient_summary}
\end{tabular} } \\
\end{table}

\subsection{TMP Data}
A homogenous TMP, namely A, of Computerized Imaging Reference Systems Inc. (CIRS), of dimension $3\times4$ cm$^2$ is used as the reference phantom. The TMP is made of zerdine, which exhibits echogenic patterns similar to those obtained from soft tissues. The speed of sound in the TMP is around $1540$ m/s. The TMP data was acquired using the same transducer settings as the \textit{in vivo} data. For the estimation of average ESD we have used two homogeneous CIRS TMPs, namely A and B, which are inclusion-free, of dimensions $3\times4$ cm$^2$ and one heterogeneous TMP, namely C, of dimensions $4.5\times4$ cm$^2$,  having a spherical inclusion of diameter $1.4$ cm. The actual average ESD, as supplied by the manufacturer, i.e, CIRS, are used as gold standards for performance evaluation of the ESD estimators.
The ESD description of the experimental phantom datasets are presented in Table \ref{Exp_dataset}.
\begin{table}[!ht]
\begin{center}
\caption{Description of experimental TMPs.}
\label{Exp_dataset}

\begin{tabular}{|c|c|c|c|}
\hline
\multirow{2}{*}{TMP Dataset} & \multirow{2}{*}{Description} & \multicolumn{2}{|c|}{\begin{tabular}[c]{@{}c@{}}Average \\   ESD ($\mu$m)\end{tabular}} \\ \cline{3-4}
                              &                              & Inclusion                                           & Background                                   \\ \hline
A                             & Homogeneous                  & -                                                  & 45                                              \\ \hline
B                             & Homogeneous                  & -                                                  & 45                                            \\ \hline
C                             & Heterogeneous                  & 70 (spherical)                                                  & 45                                               \\ \hline
\end{tabular}

\end{center}
\end{table}
\subsection{Binary Classification of Breast Lesions}
 The binary (benign-malignant) classification performances of  ESD, and the combination of ESD and MSS (both normalized) are evaluated on the dataset with the help of support vector machine (SVM), K-nearest neighbor (KNN), linear discriminant analysis (LDA), multinomial logistic regression (MNR), and Na\"ive Bayes (NB) classifiers. The best classification performance obtained has been reported in this work.  Here, we used the commonly employed ``one-versus-all'' or OVA based classification technique \cite{Moon_09}. The total data set is randomly divided into five groups for training and testing of the characterization indices as in \cite{Moon_09}. First, group $1$ is used as the test set and the remaining four groups are used as the training set for each class. Next, group $2$ is used as the test set and the remaining four groups are used as the training set. The process is repeated until the testing of each group is completed. Details on the grouping for
the binary classification are provided in Table \ref{grouping}. The classification is carried out on $159$ out of the $245$ RF data records (excluding cysts and normal breast data). The classification results include true positive (TP), true negative (TN), false positive (FP), false negative (FN), sensitivity, specificity, accuracy, positive predictive value (PPV), negative predictive value (NPV), the sum of the last five parameters, $Sum_5$, and Matthew's correlation coefficient (MCC). From TP, TN, FP, and FN, we calculate the sensitivity, specificity, accuracy, PPV, NPV and MCC \cite{Zhu_10}, \cite{Matthews_75}, \cite{Zhou_13}.  Higher values of sensitivity, specificity, accuracy, PPV, and NPV indicate better classification performance. For MCC, $+1$ indicates a perfect prediction, $0$ indicates a uniform random prediction, and $-1$ indicates an inverse prediction \cite{Matthews_75}.
\renewcommand{\arraystretch}{1.0}
\begin{table}[!ht]
\centering
\caption{Division of breast lesion data into five groups for training-testing.}
\footnotesize
\begin{tabular}{ccc}
\\
\hline
 Group no.     & Total(= Malignant+Fibro.+Inflam.)  \\
 \hline\
1         &  33 (= 12+16+5)   \\
\hline\
2         &  32 (= 11+16+5)  \\
\hline\
3          &  32 (= 11+16+5)  \\
\hline\
4         &  32 (= 11+16+5)  \\
\hline\
5         &  30 (= 11+15+4)  \\
\hline\
Total         &  159 (= 56+79+24)  \\
\hline\
 \label{grouping}
\end{tabular}
\end{table}

\subsection{Proposed ESD Estimation Technique}
The rest of this section is devoted to description of the proposed ESD and MSS estimation algorithms, which have been developed in order to produce improved breast lesion classification results. A greater focus is placed on the ESD estimation algorithm since the scheme is significantly modified from the frequency domain Gaussian form factor technique in \cite{Zachary_02}, on which the proposed ESD estimation technique is based.

\subsubsection{Preprocessing}
\begin{figure}[!ht]
\centering
\includegraphics[width=3.5in]{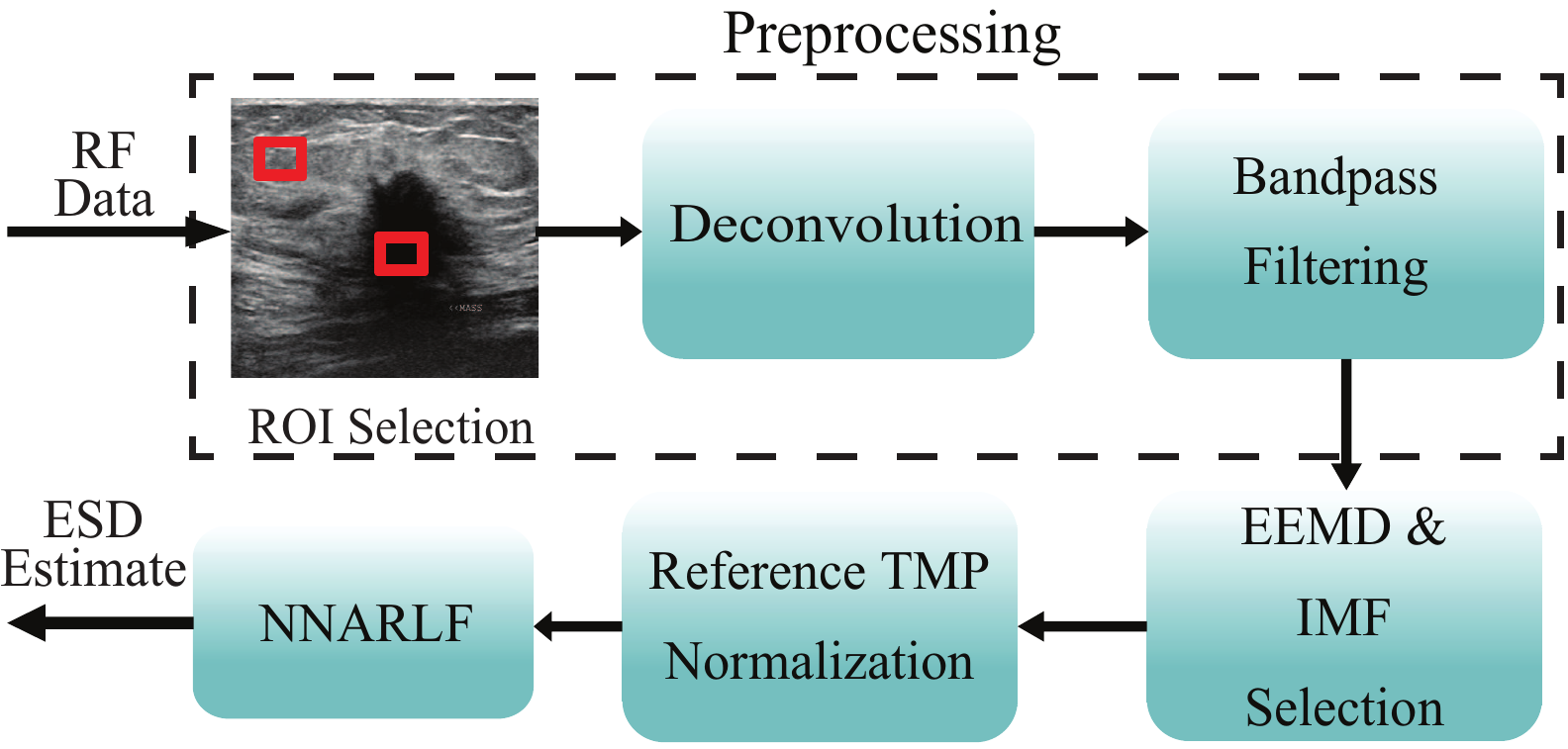}
\caption{A block diagram illustrating the proposed ESD estimation technique. NNARLF refers to nearest neighborhood average regression line fitting.}
\end{figure}
A block diagram of our proposed ESD estimation algorithm is shown in Fig. 1. The preprocessing of RF data refers to selection of regions of interest (ROI), followed by deconvolution and filtering. It is to be noted that, in order to accurately estimate ESD from the backscattered RF data, system effects, that is the impact of the system point spread function (PSF) and diffraction need to be minimized \cite{Kilmonda_16}. In this paper, a multi-step system effect minimization scheme has been proposed with deconvolution and filtering serving as the first two steps. The last step of the system effect minimization process is normalization using a reference TMP as shown in the block diagram of Fig. 1, which is discussed later in this section.

 At first, 2-D rectangular regions of interest (ROI) are selected from the B-mode images obtained from the recorded RF data. Any lesion present had been pre-identified on the B-mode image by the radiologist, who acquired the data. In case of RF data containing lesions, a suitable 2-D rectangular region within the border of the lesion is taken as the ROI. Additionally, a second ROI is selected outside the lesion to compare the ESD values inside and outside the lesion. This has been illustrated schematically in the block diagram of Fig. 1. In case of normal breast data, a suitable ROI is selected and it is ensured that the selected ROI is away from the edges of the imaging plane. The dimension of each ROI is approximately (10-12)$\times$(6.25-9.40) mm$^2$ which approximately equals to $25$ pulse lengths axially and $4$ beam widths laterally. The choice of such an ROI size has been previously shown to improve the accuracy of QUS parameter estimates \cite{Ghoshal_10}. Though ESD estimation from \textit{in vivo} data uses only one ROI, for estimating the average ESD from TMP datasets A and B, $25$ ROIs are selected, each of dimension $1\times1$ cm$^2$, while for TMP dataset C, $10$ ROIs are selected from outside the inclusion and $5$ ROIs are selected within the inclusion. In TMP dataset C, we have also selected $5$ heterogeneous ROIs across the border of the inclusion such that the ROIs encompass both the inclusion and the background. Since the manufacturers quote an average ESD values for the TMPs, selecting a large number of ROIs allows us to better evaluate the performance of the ESD estimators.

  Next, the RF data is deconvolved using a blind multi-channel least-mean squares algorithm previously proposed in \cite{Hasan_2016}. The technique in \cite{Hasan_2016} requires no prior knowledge of the PSF. It uses a modified block-based cross-correlation technique to overcome the non-stationarity of PSF and problems associated with incomplete ultrasound data acquistion. Furthermore, it employs an \textit{$l_{1}$}-norm based cost function with a damped variable step-size, which helps mitigate the impacts of noise and results in a high convergence speed. Finally, the PSF is estimated  using a regularized multiple-input/multiple-output algorithm, which is equally applicable for minimum and non-minimum phase signals. This technique has been shown to outperform most conventional deconvolution techniques \cite{Hasan_16} and is, therefore, well suited for our work, to reduce the impact of the system PSF.

Following deconvolution, an ideal bandpass filter of frequency range $2-13$ MHz is applied on the deconvolved data. It has been established in \cite{Kilmonda_16} and very recently, in \cite{Khan_19}, that the diffraction effect is most prominent in the low-frequency region of the spectrum (below $2$ MHz). It is to be noted that deconvolution shifts the backscattered signal from being centered at the pulse frequency ($10$ MHz) to lower frequencies. A upper cut-off frequency of $13$ MHz is used to eliminate any high frequency acquisition noises that may be present \cite{Khan_19}.

\subsubsection{Coherent Component Suppression using EEMD}

The backscattered RF data consist of two major components, a coherent component and a diffuse component \cite{Wear_93}. The diffuse component represents scattering from aperiodic scatterers. Although the coherent component has also been shown to be of diagnostic importance in the estimation of other QUS parameters like the MSS \cite{Bige_05}, accurate estimation of ESD requires the suppression of the coherent component \cite{Luchies_12} as the coherent component has been shown to increase the bias and variance of ESD estimates \cite{Luchies_12}. Therefore, on removal of the system PSF by  deconvolution, only the coherent and the diffuse components remain in additive form and hence, can be separated  by a suitable signal decomposition technique \cite{Nizam_17}.

Empirical mode decomposition (EMD) decomposes a signal into its IMFs in additive form \cite{Huang_98}. The main advantage of EMD over other data-driven approaches is that it requires no pre-selection of basis functions and is suitable for both stationary and non-stationary signals \cite{Huang_98}. Hence, EMD can be used to extract the diffuse component from the selected ROI from the backscattered signal after deconvolution and filtering has been carried out. However  small perturbations can  adversely effect EMD and a completely different set of IMFs may result each time it is performed. This would increase the variance of any QUS parameter estimated from these IMFs \cite{Gledhill_04}. In order to produce stable IMF estimates, EEMD is usually performed \cite{Gao_08}. In EEMD, an ensemble of random Gaussian noise is added to the backscattered signal to produce an ensemble of signals. The EMD algorithm is applied to each signal in the ensemble to produce an ensemble of IMFs. These ensemble of IMFs are then averaged to produce a new set of IMFs. The IMFs are normalized by dividing with their maximum amplitude so that no undue weight is given to any single IMF.  Selection of IMFs is a crucial issue in any algorithm involving EEMD \cite{Nizam_17}, \cite{Khan_16}, as some IMFs will contain information about the coherent scatterers while the others will contain information about the diffuse scatterers \cite{Nizam_17}. In order to identify the IMFs exhibiting diffuse scattering, a K-S test is performed. The method used is described in \cite{Georgiou_98}. The K-S classifier assumes that diffuse scatterers generate Gaussian statistics and any deviation is a result of coherent scattering. Those IMFs that show deviation from Gaussian statistics (at $5$\% significance level) are excluded and the remaining IMFs are added and used for further processing. The list of the important symbols and acronyms used in this paper are presented in Table \ref{acronym}, for convenience.

\begin{table}[!ht]
\scriptsize
\centering
\caption{Symbols and Acronyms.}
\resizebox{\columnwidth}{!}{
\begin{tabular}{ll}
\\
\hline  Symbol  &  Description\\
\hline $W(v)$  & Gated backscattered RF signal in the frequency domain\\
$T(f)$  & Combined effect of transmitted pulse and transducer sensitivity in the frequency domain\\
$D(f,z)$  & Diffraction component of $W(v)$ in the frequency domain\\
$S(f,D_{eff},n_z)$  & Scattering component of $W(v)$ in the frequency domain\\
$D_{eff}$ & Effective scatterer diameter\\
$n_z$  & Effective acoustic concentration\\
$z$  & Depth of the gated segment from the transducer face\\
$A(f,z)$& Cumulative attenuation\\
$\rho(f)$ & Attenuation coefficient (AC) in unit Nepers/cm\\
$A_c(f,z)$& Attenuation compensation function\\
$W^{'}_{comp}(v)$  & Attenuation compensated backscattered RF signal in the frequency domain\\
$w^{'}_{comp}(n)$& Attenuation compensated backscattered signal in the discrete-time domain\\
 $g_{p}(n)$ & Random Gaussian noise in the discrete-time domain\\
 $N_E$& Vector length of random Gaussian noise\\
$w^{'}p_{comp}(n)$& $w^{'}_{comp}(n)$ with 30dB Gaussian noise added, used for EEMD decomposition \\
$c_{pj}(n)$& IMFs computed from $w^{'}p_{comp}(n)$ using EEMD \\

${c_{j}}(n)$& Ensemble average of IMFs, computed over the noise vector length of $N_E$\\
$M$& No of IMFs responsible for diffuse scattering, selected through K-S test\\
$w^{'}i_{comp}(n)$ & Sum of all ensemble averaged IMF, responsible  for diffuse scattering, in the discrete-time domain\\
$W^{'}_{S,comp}(v)$& Frequency domain representation of $w^{'}i_{comp}(n)$ for sample \\
$W^{'}_{R,comp}(v)$& Frequency domain representation of $w^{'}i_{comp}(n)$ for reference \\
RF & Radio-frequency \\
ESD & Effective scatterer diameter  \\
EAC & Effective acoustic concentration\\
MSS & Mean scatterer spacing\\
AR & Autoregressive \\
QUS &  Quantitative ultrasound\\
AC & Attenuation coefficient\\
TMP & Tissue-mimicking phantom\\
CIRS & Computerized Imaging Reference Systems Inc. \\
ROI & Region of interest\\
PSF & Point spread function\\
EEMD & Ensemble empirical mode decomposition\\
IMF & Intrinsic mode function\\
K-S test & Kolmogorov-Smirnov test \\
SD & Standard deviation \\
MASD & Minimization of the average squared deviation \\
MAPE & Mean absolute percentage error\\
NN & Nearest neighborhood \\
NNARLF & Nearest neighborhood average regression line fitting\\
SAC & Spectral autocorrelation\\
SSA & Singular spectrum analysis\\
MCC & Matthew's correlation coefficient\\
ROC & Receiver operating characteristic\\
\hline
\label{acronym}
\end{tabular}} \\
\end{table}

\subsubsection{NNARLF for ESD Estimation}

The gated backscattered RF signal intensity, $W(v)$, where $v = \{f,D_{eff},z,n_z\}$ is the set of variables on which the backscattered RF signal depends, at the transducer face can be expressed in the frequency domain as \cite{Kim_07}
\begin{equation}
\label{3_BS_RF} W(v) = T(f) \cdot D(f,z) \cdot A(f,z) \cdot S(f,D_{eff},n_z),
\end{equation}
where $T(f)$ represents the combined effect of the transmit pulse and the transducer sensitivity (electro-acoustic and acousto-electric transfer functions); $D(f,z)$ is the effect of diffraction; $A(f,z)$ is the cumulative attenuation in the soft tissue and $S(f,D_{eff},n_z)$ represents the scattering properties of the tissue, including the ESD $(D_{eff})$ and EAC $(n_z)$; $z$ being the depth of the gated segment from the transducer face \cite{Kim_07}, \cite{Labyed_11}.
We acquire RF signals from the tissue sample and the reference TMP. The backscattered RF signal is then deconvolved and filtered, as the first two steps of the system effect minimization process. The resulting signal, in the frequency domain, is given by $W^{'}(v)$.
 The cumulative attenuation, $A(f,z)$, in soft tissues is a function of frequency $f$ and depth $z$, and can be expressed as \cite{Kuc_85}
\begin{eqnarray}
\label{A} A(f,z)=e^{-4\rho(f)z}=10^{-2\rho(f)z/10},
\end{eqnarray}
where $\rho(f)$ denotes the attenuation coefficient (AC) in unit Nepers/cm. It is reported in \cite{Kuc_85} that $\rho(f)$ demonstrates a linear frequency dependence. Therefore, it can be written as $\rho(f)=\beta\cdot f$, where $\beta$ denotes the AC in Nepers/cm/MHz. By compensating for the effect of frequency dependent attenuation, using a depth-dependent (dependent on $z$) spectral average based attenuation estimation technique developed in \cite{Hasan_2013}, we get the compensated backscattered RF signal in the frequency domain as
\begin{eqnarray}
\label{3_BS_RF_cmp} \nonumber W^{'}_{comp}(v) &=& W^{'}(v)A_c(f,z) \\ &=& T^{'}(f) \cdot D^{'}(f,z) \cdot S(f,D_{eff},n_z),
\end{eqnarray}
where $A_c(f,z)$ is the frequency dependent attenuation compensation function defined as
\begin{equation}
\label{3_atn_cmp} A_c(f,z) = A^{-1}(f,z);
\end{equation}
and $T^{'}(f)$ and $D^{'}(f,z)$ are residual effects of the system PSF and diffraction, respectively, remaining after deconvolution and filtering.

Now, let us consider the attenuation-compensated signal in the discrete-time domain to be $w^{'}_{comp}(n)$. To perform EEMD of the attenuation-compensated RF data, an ensemble of $N_E$ random Gaussian noise, $g_{p}(n)$ ($p$=1,$\cdots$, $N_E$) is added to $w^{'}_{comp}(n)$. That is, the ensemble is given by \cite{Huang_05}
 \begin{equation}
 w^{'}p_{comp}(n)=w^{'}_{comp}(n)+g_{p}(n), p=1,\cdots, N_E.
 \end{equation}
 The signal-to-noise ratio between the RF data and the Gaussian noise is kept at $30$ dB.  After that, the EMD algorithm \cite{Huang_98} is applied  to each of the signals in the ensemble to decompose them into a sum of their IMFs, $c_{pj}(n)$, $j$=1,$\cdots$, $K$, where $K$ is the number of IMFs and a residue, $r_{p}(n)$, given by \cite{Huang_05}
 \begin{equation}
 w^{'}p_{comp}(n)=\sum_{j=1}^{K}c_{pj}(n)+r_{p}(n), p=1,\cdots, N_E.
 \end{equation}
Finally, the IMFs using EEMD are obtained from the ensemble average \cite{Huang_05}
\begin{equation}
\overline{c_{j}}(n)=\frac{1}{N_E}\sum_{p=1}^{N_E}c_{pj}(n), j=1,\cdots, K.
\end{equation}
The IMFs responsible for diffuse scattering, $c_{d}(n)$, $d$=1,$\cdots$, $M$, where $M$ is the number of IMFs (normalized) responsible for diffuse scattering, are then identified by the K-S test \cite{Georgiou_98}. The IMFs are normalized using their amplitude. Normalization of the IMFs is done to ensure that no undue weight is given to any one of the IMFs. The attenuation-compensated RF data, for both the sample and the TMP, are then replaced by the summation of the IMFs responsible for diffuse scattering as
\begin{equation}
w^{'}i_{comp}(n)=\sum_{d=1}^{M}c_{d}(n), d=1,\cdots, M,
\end{equation}
where, $w^{'}i_{comp}(n)$ is the IMF-replaced signal in the discrete-time domain.

If we consider $W^{'}_{S,comp}(v)$ and $W^{'}_{R,comp}(v)$ as the IMF-replaced signals in the frequency domain for the sample and the reference, respectively, for the same average sound speed in the reference and sample tissues, the residual diffraction terms can be considered equal. The sound speed in the reference is known from the manufacturer's specifications to be approximately equal to that of soft tissues. Moreover, since the sample and reference data are obtained using the same transducer settings, the residual effect of the system PSF can also be considered equal in the sample and the reference. It is to be noted that the reference data is also passed through the same preprocessing steps as the sample data. Hence, on dividing $W^{'}_{S,comp}(v)$ by $W^{'}_{R,comp}(v)$, as the final step of the multi-step system effect minimization process, we get the normalized spectra, $W_{norm}(v)$, as
\begin{eqnarray}
\label{3_Norm_cmp_ps} W_{norm}(v) = \frac{W^{'}_{S,comp}(v)}{W^{'}_{R,comp}(v)} = \frac{S_S(f,D_{eff,S},n_{z,S})}{S_R(f,D_{eff,R},n_{z,R})},
\end{eqnarray}
where  $S_R(f,D_{eff,R},n_{z,R})$  and $S_S(f,D_{eff,S},n_{z,S})$ represents the scattering properties of the sample and reference, respectively.
Taking the logarithm on both sides of (\ref{3_Norm_cmp_ps}), we get
\begin{equation}
\label{3_log_Norm_cmp_ps} 10\log{W_{norm}(v)} = 10\log{\frac{S_S(f,D_{eff,S},n_{z,S})}{S_R(f,D_{eff,R},n_{z,R})}}.
\end{equation}

A model for tissue scattering, $S(f,D_eff,n_z)$,in the frequency domain was developed in \cite{Lizzi_97} for clinical array systems given by
\begin{eqnarray}
\label{3_TS} \nonumber S(f,D_{eff},n_z) = \frac{185Lq^2(\frac{D_{eff}}{2})^6n_zf^4}{[1+2.66(fq(\frac{D_{eff}}{2})^2)]}\times \\ e^{-12.159f^2(\frac{D_{eff}}{2})^2}
\end{eqnarray}
with $L$ the gate length (mm), $q$ the ratio of aperture radius to distance from the region of interest, $f$ the frequency in MHz, and $D_{eff}$ the ESD in mm. The model was derived using a Gaussian form factor model and it has been previously established that a Gaussian form factor is suitable for modeling  the scattering from human tissues \cite{Oleze_02}, \cite{Insana_90_JASA}. The quantity, $n_z$, is termed the EAC and is defined in \cite{Oleze_02}. For sample and reference, the tissue scattering is then represented as
\begin{eqnarray}
\label{3_TS_S}  \nonumber  S_S(f,D_{eff,S},n_{z,S}) = \frac{185Lq^2(\frac{D_{eff,S}}{2})^6n_{z,S}f^4}{[1+2.66(fq(\frac{D_{eff,S}}{2})^2)]} \times \\ e^{-12.159f^2(\frac{D_{eff,S}}{2})^2},\\
\label{3_TS_R} \nonumber S_R(f,D_{eff,R},n_{z,R}) = \frac{185Lq^2(\frac{D_{eff,R}}{2})^6n_{z,R}f^4}{[1+2.66(fq(\frac{D_{eff,R}}{2})^2)]} \times\\  e^{-12.159f^2(\frac{D_{eff,R}}{2})^2}.
\end{eqnarray}
Here, $D_{eff,S}$ is the ESD to be estimated from the sample (tissue) and $D_{eff,R}$ is the ESD of the reference (TMP), which is known from the manufacturer's specifications. Dividing (\ref{3_TS_S}) by (\ref{3_TS_R}), and considering the fact that $2.66(fq\frac{D_{eff}}{2})^2\ll1$ \cite{Zachary_02}, we get the normalized tissue scattering as
\begin{equation}
\label{3_TS_S2R} \frac{S_S(f,D_{eff,S},n_{z,S})}{S_R(f,D_{eff,R},n_{z,R})} = \frac{D_{eff,S}^6 n_{z,S}}{D_{eff,R}^6 n_{z,R}}e^{-3.03975(D_{eff,S}^2-D_{eff,R}^2)f^2}.
\end{equation}
 In order to estimate the ESD from breast tissues, TMP dataset A has been used as the reference. In case of estimating the ESD from the TMP datasets, TMP dataset A has been used as reference for TMP datasets B and C and TMP dataset B has been used as the reference for TMP dataset A.
Next, taking logarithm on both sides of (\ref{3_TS_S2R}) yields
\begin{eqnarray}
\label{3_log_TS_S2R} \nonumber 10\log{\frac{S_S(f,D_{eff,S},n_{z,S})}{S_R(f,D_{eff,R},n_{z,R})}} = 10\log{\frac{D_{eff,S}^6 n_{z,S}}{D_{eff,R}^6 n_{z,R}}} -13.20\times \\ (D_{eff,S}^2-D_{eff,R}^2) f^2.
\end{eqnarray}
To estimate ESD, we fit a regression line through the usable (i.e., $-6$~dB) bandwidth of the normalized log scattering power spectrum. From (\ref{3_log_Norm_cmp_ps}), this is equivalent to fitting a line through the usable bandwidth of the normalized log power spectrum.
Assuming $f^2 = x$, the regression line can be expressed as
\begin{equation}
\label{3_str_ln} y = mx + c
\end{equation}
with
\begin{eqnarray}
\label{3_def_y} y &=& 10\log{\frac{S_S(f,D_{eff,S},n_{z,S})}{S_R(f,D_{eff,R},n_{z,R})}},\\
\label{3_def_m} m &=& -13.20\times(D_{eff,S}^2-D_{eff,R}^2),\\
\label{3_def_c} c &=& 10\log{\frac{D_{eff,S}^6 n_{z,S}}{D_{eff,R}^6 n_{z,R}}}.
\end{eqnarray}

Using average block power spectra generated from spatial signal blocks of sufficiently large size, ESD can be estimated from (\ref{3_def_m}). On the other hand, probability that a single gated RF block includes regions of heterogenous tissues increases with the block size. Hence to trade-off between homogeneity in the spatial blocks and consistency of the estimated power spectrum, we use an NNARLF algorithm. We assume that the ESD of the neighboring tissues of the scattering particles in the neighboring blocks are almost the same for their physical proximity. We later generate 2-D ESD colormaps on a B-mode image and also carry out experiments on the TMP and \textit{in vivo} data to show the advantages of employing such a weighted exponential neighborhood. Therefore, we calculate an average regression line as the weighted average of the regression lines of the neighboring blocks as
\begin{eqnarray}
\label{3_nn_str_ln} Y(i_s,j_s) = \frac{\sum_{i_0=i_s-L_a}^{i_s+L_a}\sum_{j_0=j_s-L_l}^{j_s+L_l}y(i_0,j_0)w^{(i_s.j_s)}(i_0,j_0)}{\sum_{i_0=i_s-L_a}^{i_s+L_a}\sum_{j_0=j_s-L_l}^{j_s+L_l}w^{(i_s.j_s)}(i_0,j_0)}
\end{eqnarray}
where $Y(i_s,j_s)$ denotes the weighted average value of $y(i_s,j_s)$, and $w^{(i_s.j_s)}(i_0,j_0)$ is the exponential weight function for an interrogated point $(i_s,j_s)$ on the 2-D ESD map, defined as
\begin{eqnarray}
\label{3_wgt_fn} w^{(i_s,j_s)}(i_0,j_0) = e^{-|\lambda_a(i_0-i_s)|-|\lambda_l(j_0-j_s)|},
\end{eqnarray}
where $\lambda_a$ and $\lambda_l$ denote the weighting factors in the axial and lateral directions, respectively, and $i_s-L_a\leq i_0\leq i_s+L_a$ and $j_s-L_l\leq j_0\leq j_s+L_l$. $L_a$ and $L_l$ are the nearest neighborhood (NN) factors in the axial and lateral directions, respectively. From (\ref{3_wgt_fn}) it is evident that $w^{(i_s.j_s)}(i_0,j_0)$ has the maximum value (unity) at $(i_0,j_0) = (i_s,j_s)$. We define $w^{(i_s,j_s)}$ in a way such that in the averaging process, the logarithm of measured power spectrum of a neighboring window is properly weighted to have less contribution with increasing distance from the interrogated block. A 2-D weighted exponential neighborhood having $L_a=L_I=5$ is illustrated in Fig. 2. The values on the weight axis are arbitrary but show an exponential decay as we move away from the interrogated window both axially and laterally.

\begin{figure} [!ht]
\includegraphics[width=3in]{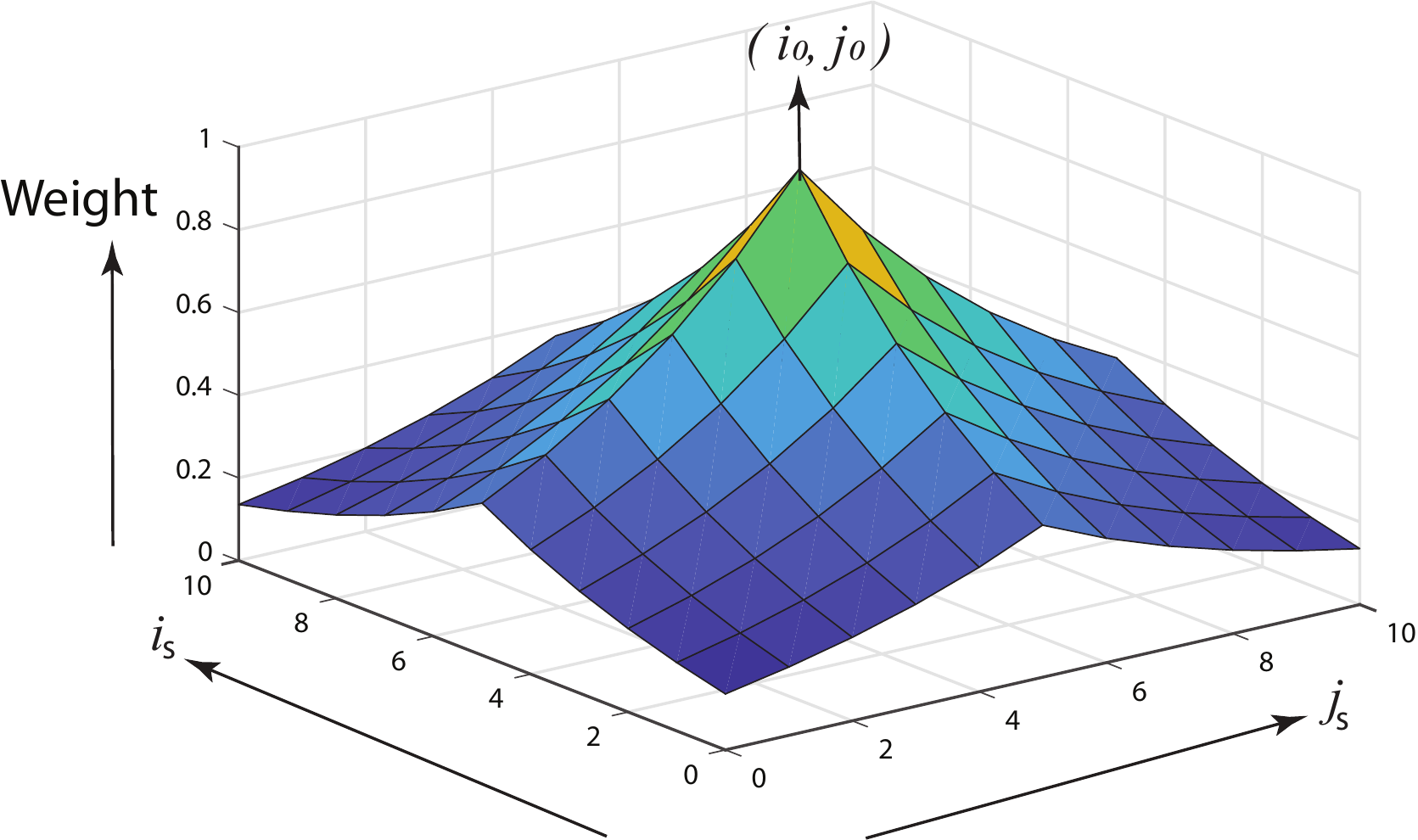}
\caption{An illustration of the exponentially weighted neighborhood.}
\end{figure}
Substituting the value of $y$ from (\ref{3_str_ln}) into (\ref{3_nn_str_ln}), we get
\begin{eqnarray}
\centering
\label{3_nn_str_ln1}  Y(i_s,j_s) = \frac{\sum_{i_0=i_s-L_a}^{i_s+L_a}\sum_{j_0=j_s-L_l}^{j_s+L_l}(m(i_0,j_0)+c(i_0,j_0))w^{(i_s.j_s)}(i_0,j_0)}{\sum_{i_0=i_s-L_a}^{i_s+L_a}\sum_{j_0=j_s-L_l}^{j_s+L_l}w^{(i_s.j_s)}(i_0,j_0)}.
\end{eqnarray}

If we define the weighted average value of the slope as
\begin{equation}
\label{3_w_slope} M(i_s,j_s) = \frac{\sum_{i_0=i_s-L_a}^{i_s+L_a}\sum_{j_0=j_s-L_l}^{j_s+L_l}m(i_0,j_0)w^{(i_s.j_s)}(i_0,j_0)}{\sum_{i_0=i_s-L_a}^{i_s+L_a}\sum_{j_0=j_s-L_l}^{j_s+L_l}w^{(i_s.j_s)}(i_0,j_0)},
\end{equation}
and the weighted average value of the intercept as
\begin{equation}
\label{3_w_slope} C(i_s,j_s) = \frac{\sum_{i_0=i_s-L_a}^{i_s+L_a}\sum_{j_0=j_s-L_l}^{j_s+L_l}c(i_0,j_0)w^{(i_s.j_s)}(i_0,j_0)}{\sum_{i_0=i_s-L_a}^{i_s+L_a}\sum_{j_0=j_s-L_l}^{j_s+L_l}w^{(i_s.j_s)}(i_0,j_0)},
\end{equation}
then (\ref{3_nn_str_ln1}) can be written in the form of a regression line given by
\begin{equation}
\label{3_avg_str_ln} Y(i_s,j_s) = M(i_s,j_s)x+C(i_s,j_s).
\end{equation}
Now, from the slope, $M$, of the regression line that fits (\ref{3_avg_str_ln}), we can estimate the ESD, $D_{eff,S}$, at the point $(i_s,j_s)$ using (\ref{3_def_m}) as
\begin{equation}
\label{3_avg_D}D_{eff,S} = \sqrt{-\frac{M(i_s,j_s)}{13.20}+D_{eff,R}^{2}}.
\end{equation}
\newline
In order to estimate the block power spectra of an interrogated block with higher resolution, the block is divided into $1$-D segments with consecutive window segments in a block having 50\% axial overlapping. The windowed segments are gated by the Hamming window. The block power spectra are calculated using the Welch method \cite{Welch_67}.  We select $L_a$ = $L_l$= 5 as the NN factors to estimate the local ESD for a particular interrogated block. The impact of varying the NN factors on the ESD estimation is discussed in the results section.

\subsection{MSS Estimation Technique}
Another QUS parameter which has previously been successfully used for breast tissue characterization is MSS  \cite{Bige_05}, \cite{Tadayyon_14}, \cite{Nizam_17}.  Specifically, a MSS estimation algorithm developed in \cite{Nizam_17} produced promising binary classification results, albeit, on a small dataset. MSS estimation on the same dataset as ESD allows us to use multiple QUS parameters for breast lesion classification. The use of multiple QUS parameters for tissue characterization has been previously shown to produced improved results \cite{Tadayyon_14}, \cite{Golub_93}, \cite{Ara_2017}. What is more, because the MSS is estimated from the coherent component of the backscattered RF data and EEMD separates the diffuse and coherent components, the MSS may be estimated simultaneously with the ESD, using a branching scheme. In the technique proposed in \cite{Nizam_17}, the AR spectrum was estimated from the IMFs produced by EEMD of the deconvolved RF data. In that work, the IMFs, responsible for coherent scattering (from which the MSS is estimated), were identified using an empirical criterion. The K-S test was only used to validate whether or not the selected IMFs are indeed a source of coherent scattering. In this work, we have ameliorated the previously proposed MSS estimation technique by incorporating the automatic IMF selection scheme, along with bandpass filtering of the deconvolved RF data. A block diagram to illustrate the MSS estimation technique is shown in Fig. 3. We present the values of the average MSS estimates for different types of breast tissues and the performance of the MSS estimator as a biomarker for breast lesion classification in the results section.

\begin{figure}[!ht]
\includegraphics[width=3.5in]{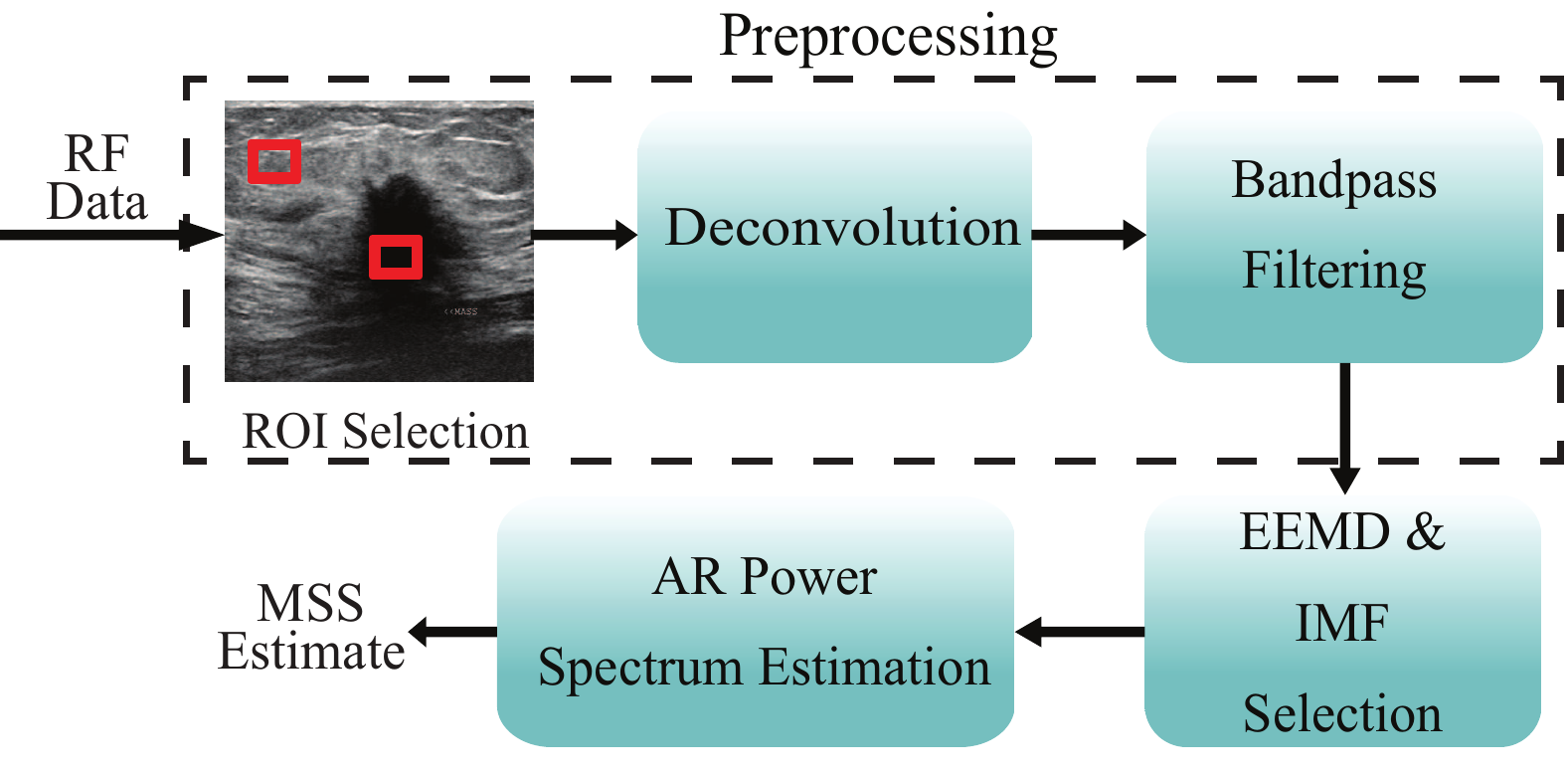}\DeclareGraphicsExtensions{.eps}
\caption{A block diagram illustrating the proposed MSS estimation technique.}
\end{figure}

\section{Results}

In this section, we present the classification results obtained by combining ESD with MSS. We also present the single-parameter classification results obtained using our ESD estimates and compare that classification performance with the classification results obtained using ESD estimated from other techniques that employ a Gaussian form factor model \cite{Oleze_02}, \cite{Zachary_02}. In addition, we show the single-parameter classification performance of the MSS estimator and compare its performance with some existing techniques for MSS estimation \cite{Tadayyon_14}, \cite{Pereira_01}. However, at first, we focus on the ESD estimates obtained using our proposed method and validate the accuracy of our proposed algorithm by estimating the ESD from TMPs. Additionally, we present the results of ESD estimated using some of the conventional Gaussian form factor based techniques \cite{Oleze_02}, \cite{Zachary_02}. Moreover, we present numerical results to validate the different steps of the proposed ESD estimation algorithm and hence, establish the suitability of the proposed ESD estimator for breast lesion classification. Furthermore, we present the MSS estimation results on \textit{in vivo} breast tissues and compare the estimates with some previously proposed techniques \cite{Tadayyon_14}, \cite{Pereira_01}.

\subsection{ESD Estimation Results on Tissue-mimicking Phantoms}
To check the accuracy of our proposed average ESD estimation technique and thus, justify the use of these ESD estimates for breast lesion classification, we use three CIRS experimental TMPs for which the average ESD values in the inclusion and background are available from the manufacturer. The average ESD of these phantoms are also estimated using the MASD based method \cite{Oleze_02}, and the frequency domain method \cite{Zachary_02}. It is to be noted that a Faran form factor is usually employed for modeling the scattering from TMPs \cite{Fran_12}. But, since tissue scattering is more accurately modeled by a Gaussian form factor, to apply (\ref{3_Norm_cmp_ps}), a Gaussian form factor model has been applied for the TMPs as well. Moreover, Gaussian form factor models have also been previously used to model the scattering from TMPs \cite{Mamou_13}. Table \ref{TMP} presents the actual average ESD and the average ESD estimated using our proposed algorithm as well as the other techniques for the experimental phantoms. The average ESD represents the average of the ESD  values estimated from each of the ROIs within a particular TMP. It is evident that our proposed algorithm estimates the ESD for all three TMP data with a higher degree of accuracy compared to the other methods as reflected by a lower mean absolute percentage error (MAPE) value. Moreover, our method also shows a lower SD of estimates compared to the other methods.
\begin{table*}[!ht]
\begin{center}
\caption{Estimated ESD values (in $\mu$m) with SD (in bracket) from experimental TMPs.}
\label{TMP}
\renewcommand{\arraystretch}{0.5}
\resizebox{\textwidth}{!}{
\begin{tabular}{|l|l|l|l|l|l|}
\hline
Method                                  & TMP  & Average ESD  & MAPE  & Average ESD  & MAPE \\
 & Dataset & of Background& ($\%$)& inside Inclusion & ($\%$)\\
  & Dataset & ($\mu$m) ($\pm$SD)& ($\%$)&  ($\mu$m) ($\pm$SD) & \\ \hline
\multirow{3}{*}{MASD \cite{Oleze_02}}      & A            & 54.82($\pm8.88$)                               & 21.82           & -                                      & -                 \\
                                        & B            & 52.18($\pm8.09$)                                & 15.96           & -                                      &-                  \\
                                        & C            & 50.91($\pm7.99$)                                & 13.13           & 74.91($\pm7.87$)                                   & 7.01             \\ \hline

\multirow{3}{*}{Frequency Domain \cite{Zachary_02}} & A            & 53.84($\pm7.11$)                                & 19.64           & -                                      &                  \\
                                        & B            & 50.10($\pm7.07$)                                & 11.33           & -                                      &-                  \\
                                        & C            & 48.97($\pm7.39$)                                & 8.82           & 73.61($\pm7.09$)                                   & 6.71            \\ \hline

\multirow{3}{*}{Proposed}        & A            & 47.02($\pm5.89$)                                & 4.49            & -                                      &                  \\
                                        & B            & 47.76($\pm6.01$)                                & 6.13            & -                                      &-                  \\
                                        & C            & 47.78($\pm5.85$)                                & 6.18            & 73.02($\pm6.33$)                                   & 5.15   \\ \hline

\end{tabular}
}
\end{center}
\end{table*}

We have also estimated the ESD from the heterogeneous ROIs of TMP C using our proposed method. A scatter plot showing how the ESD varies as we move laterally from a region just outside the inclusion across the border of the inclusion to a region within the inclusion is presented in Fig. 4. It is evident from the figure that our proposed method is able to reliably estimate the ESD in the homogeneous regions inside and outside the inclusion with a mean value of approximately $48$ $\mu$m outside the inclusion (represented by a red line) and a mean value of approximately $74$ $\mu$m within the inclusion (represented by a green line). There is a sharp change in the average ESD values across the border of the inclusion, as expected.
\begin{figure}[!ht]
\centering
\includegraphics[width=3.5in]{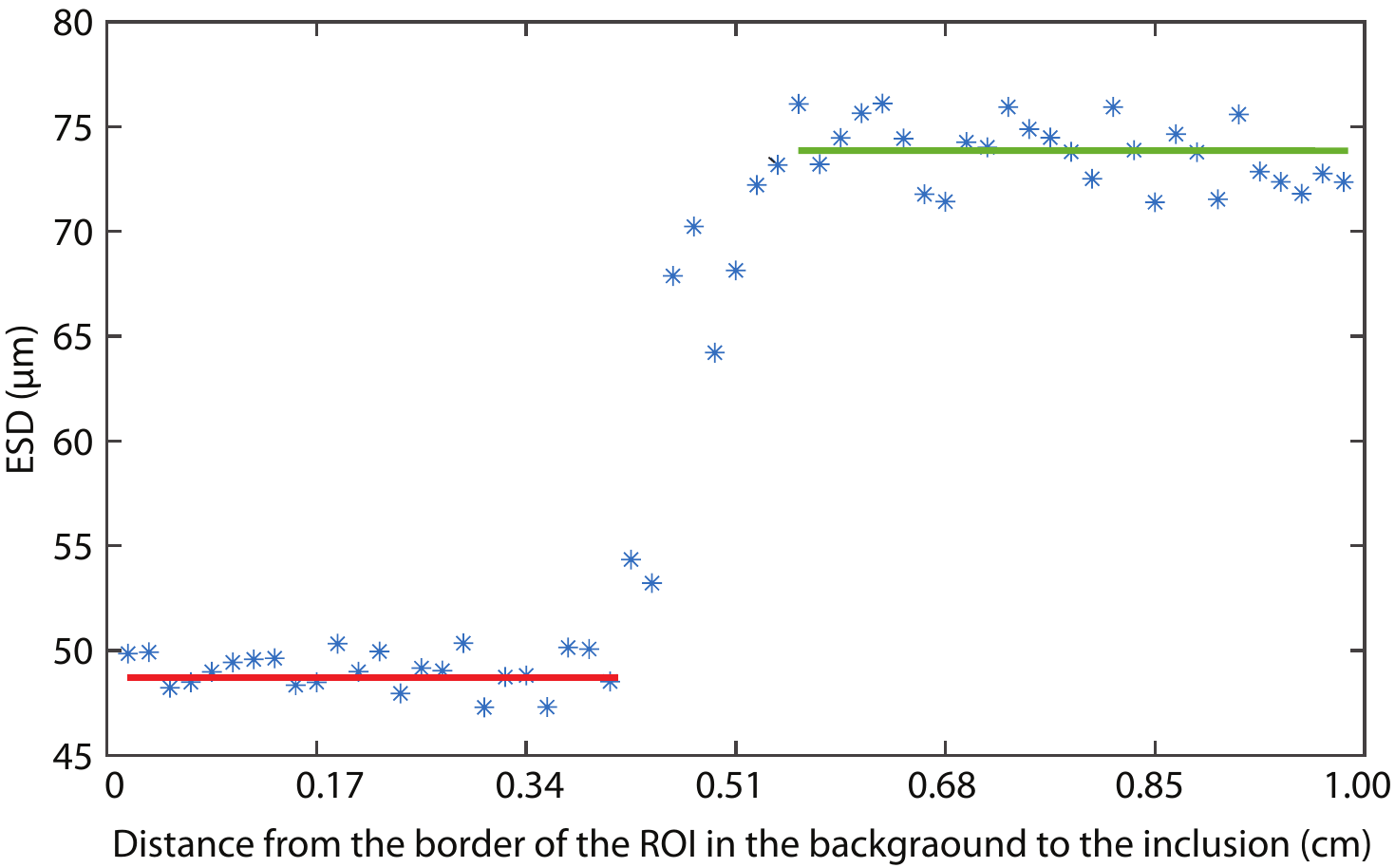}
\caption{Scatter plot showing the variation of ESD across different windows on moving laterally from the edge of the ROI in the TMP background to the edge of the ROI within the inclusion.}
\end{figure}
Now, in order to study how the different steps in the proposed algorithm impacts the accuracy of the ESD estimates, the ESD was estimated for  several ROIs of TMP datasets A, B, and C by removing the different steps shown in the flow chart of Fig. 1 one by one while retaining the others, that is, an ablation approach is taken. The results are presented in Fig. 5 in the form of a bar plot.  The first column of the bar plot presents the performance of the proposed algorithm on the same datasets for comparison. It can be seen that the overall estimation accuracy is most adversely impacted on removing the EEMD step. There is also a slight rise in the SD of estimates on removing the EEMD step. Removal of the any one of the system effect minimization steps, i.e, deconvolution, filtering, and normalization using a reference TMP, also noticeably impacts the ESD estimation accuracy, with a slight increase in the SD of ESD estimates in each case. Hence, it justifies the use of a multi-step system effect minimization technique since removal of any one of these steps has detrimental effect on the overall ESD estimation accuracy. The weighted neighborhood step seems to have the least impact on the overall accuracy. However, this step is seen to have a more negative impact on the SD of the estimates compared to the other steps of the proposed method.

\begin{figure}[!ht]
\centering
\includegraphics[width=3.5in]{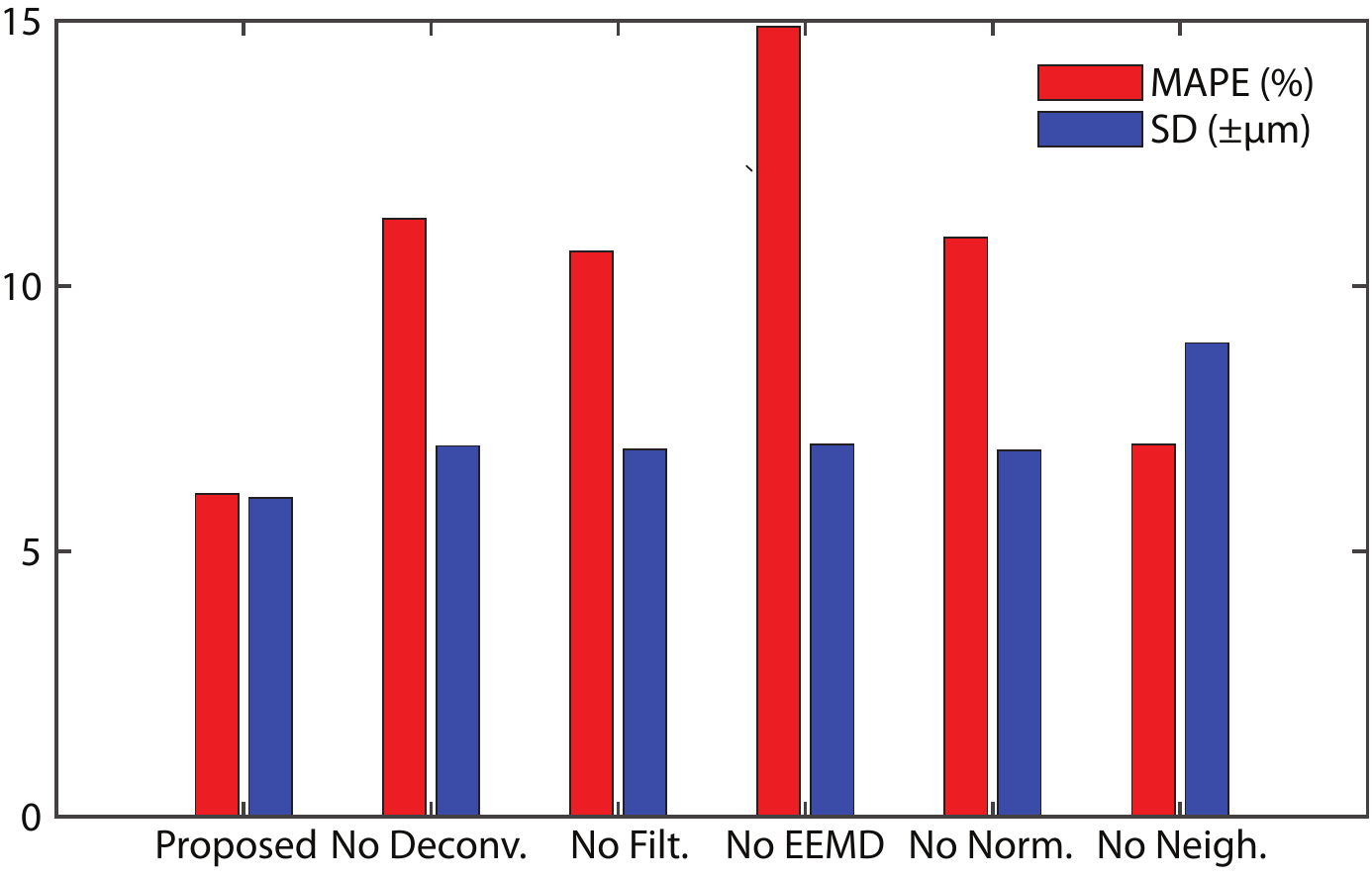}
\caption{Bar plot showing the impact on the percentage error and SD of the ESD estimates from the experimental TMPs on removing the different steps from our proposed algorithm.}
\end{figure}
\subsection{ESD and MSS Estimation Results on \textit{In Vivo} Breast Tissues}
 The results of estimating the ESD of \textit{in vivo} breast tissues using various techniques are presented in Table \ref{ESD_result}. It is evident that, using our proposed method, the estimated ESD values for fibroadenoma and malignant breast tissues fall within the range of ESD values previously reported in the literature \cite{Tadayyon_14}. According to the best of our knowledge, the ESD values of inflammatory lesions and cystic lesions have not been previously reported. The ESD values in the region outside the lesions are consistent in the range between $70-80$ $\mu$m, which is similar to normal tissues.  The ESD estimates for inflammatory tissues show little or no deviation from this range. The estimated ESD values of cysts are rather erratic (that is, the proposed technique often fails to produce any ESD estimates) and show a high SD and hence, are not presented. This is true for all the techniques used to obtain the ESD estimates in Table \ref{ESD_result}. Therefore, it can be concluded that ESD estimates of cystic lesions are of no diagnostic importance. This is in good concordance with the anatomy of the cysts as they are fluid-filled sacs and thus, scattering from cysts will largely be absent and some inconsistent scattering may occur due to debris (such as those present in complex cysts) \cite{Berg_10}. Hence, they are excluded from the dataset when benign-malignant classification of breast lesions is carried out. Furthermore, it is seen that the average ESD value of malignant lesions is greater than that of benign lesions which also conforms with the previously reported results \cite{Tadayyon_14}. The estimated average ESD values using the MASD-based method \cite{Oleze_02} and the original frequency domain method \cite{Zachary_02} are also presented in Table \ref{ESD_result}. We see that the SD of ESD estimates are significantly higher than that of our proposed method and the separation between the average ESD values for malignant lesions and fibroadenomas are also smaller.

 \begin{table*}
\centering
\caption{Estimated values of ESD with SD (in bracket) using different methods for normal and pathological tissues.}
\label{ESD_result}
\renewcommand{\arraystretch}{1.5}
\resizebox{\textwidth}{!}{

\centering
\begin{tabular}{|l|l|l|l|l|l|l|l|l|}
\hline
\multicolumn{8}{|l|}{\hspace{8cm}ESD ($\pm$SD) ($\mu$m)}                                                                                                                          \\ \hline
         & \multicolumn{2}{l|}{\hspace{1cm}Malignant}                        & \multicolumn{2}{l|}{\hspace{0.80cm}Fibroadenoma}    & \multicolumn{2}{l|}{\hspace{0.90cm}Inflammatory}  &  {Normal}     \\ \hline
Method & \multicolumn{1}{l|}{\hspace{0.75cm}in }              &\hspace{0.30cm} out            & \hspace{0.75cm}in              & \hspace{0.50cm}out           & \hspace{0.75cm}in               & \hspace{0.50cm}out  &           \\ \hline
 MASD \cite{Oleze_02}& \multicolumn{1}{l|}{100.67} & 86.12& 94.09 & 82.10  & 88.31 & 90.35& 76.13 \\
& \multicolumn{1}{l|}{($\pm$21.13)} &  ($\pm$8.41) & ($\pm$18.17) &  ($\pm$7.24) & ($\pm$11.34) & ($\pm$10.01) & ($\pm$7.01) \\ \hline
Frequency domain \cite{Zachary_02}   & \multicolumn{1}{l|}{109.21}  & 80.04  & 101.41 & 79.31  & 81.24  & 82.45  & 75.88 \\
& \multicolumn{1}{l|}{($\pm$17.34)} & ($\pm$8.21) &  ($\pm$13.12) &  ($\pm$8.54) &  ($\pm$9.81) &  ($\pm$11.01) & ($\pm$ 6.74) \\ \hline
Proposed & \multicolumn{1}{l|}{123.05 } & 74.90  & 98.71  & 74.89  & 75.72  & 75.77 & 75.12  \\
 & \multicolumn{1}{l|}{($\pm$8.85)} & ($\pm$4.19) &  ($\pm$9.55) & ($\pm$4.11) &  ($\pm$4.09) &  ($\pm$4.07)&  ($\pm$4.01) \\\hline
\end{tabular}
}
\end{table*}

An important factor that has to be taken into consideration while estimating ESD from \textit{in vivo} tissues is the impact of the kernel size of the weighted exponential neighborhood. The algorithm involves the use of a $5\times5$ weighted exponential neighborhood. The use of such a neighborhood allows the modeling of tissue homogeneity over a small region rather than a large spatial block where the tissue becomes more heterogeneous. To  investigate the impact of the size of the neighborhood on the ESD estimates, the ESD is estimated using our proposed algorithm for normal tissues for no neighborhood, a neighborhood of size $3\times3$, and a neighborhood of size $8\times8$. It has already been stated that the original results are produced for a neighborhood of size $5\times5$.  The results are shown in Table \ref{kernel}. It is clear from the table that the choice of neighborhood size does not greatly effect the average value of the ESD estimates. However, the SD of estimates seem to decrease significantly for a neighborhood size close to our selected one. A large neighborhood size or no neighborhood increases the SD of estimates and this could be attributed to the increased heterogeneity for a large spatial block of tissue. This observation is consistent to that obtained for experimental TMPs where removal of the weighted neighborhood step adversely impacted the SD of estimates.
\begin{table}[!h]
\centering
\small
\caption{Estimated values of ESD with SD (in bracket) for normal tissues with different neighborhood sizes.}
\label{kernel}
\begin{tabular}{ll}
\hline
Size of Neighborhood & ESD ($\pm$ SD) ($\mu$m)\\ \hline
No Neighborhood & 76.89 ($\pm$ 8.56)\\
$3\times3$ & 74.99 ($\pm$ 4.09)\\
$5\times5$  & 75.12 ($\pm$ 4.01)\\
$8\times8$  & 76.46 ($\pm$ 7.56)\\ \hline

\end{tabular}
\end{table}
To further substantiate our argument for choosing an exponentially weighted neighborhood, we produce, in Fig. 6, an ESD map for a representative fibroadenoma tissue for two ROIs taken within the border of the lesion and two ROIs taken outside the lesion. It is clear that the ESD values are consistent (having a low SD) across the ROIs (both inside and outside the lesion) and hence, a weighted exponential neighborhood is a good model of tissue structure. The step-wise ablation technique employed for the ESD estimation from experimental TMPs cannot be adopted for the ESD estimation from \textit{in vivo} tissues since the gold standard (i.e., histopathology) for all the data records are not available. However, the ablation approach has been applied for the classification results based on ESD estimates from \textit{in vivo} tissues in the classification results subsection.

\begin{figure}[!ht]
\centering
\includegraphics[width=3.5in]{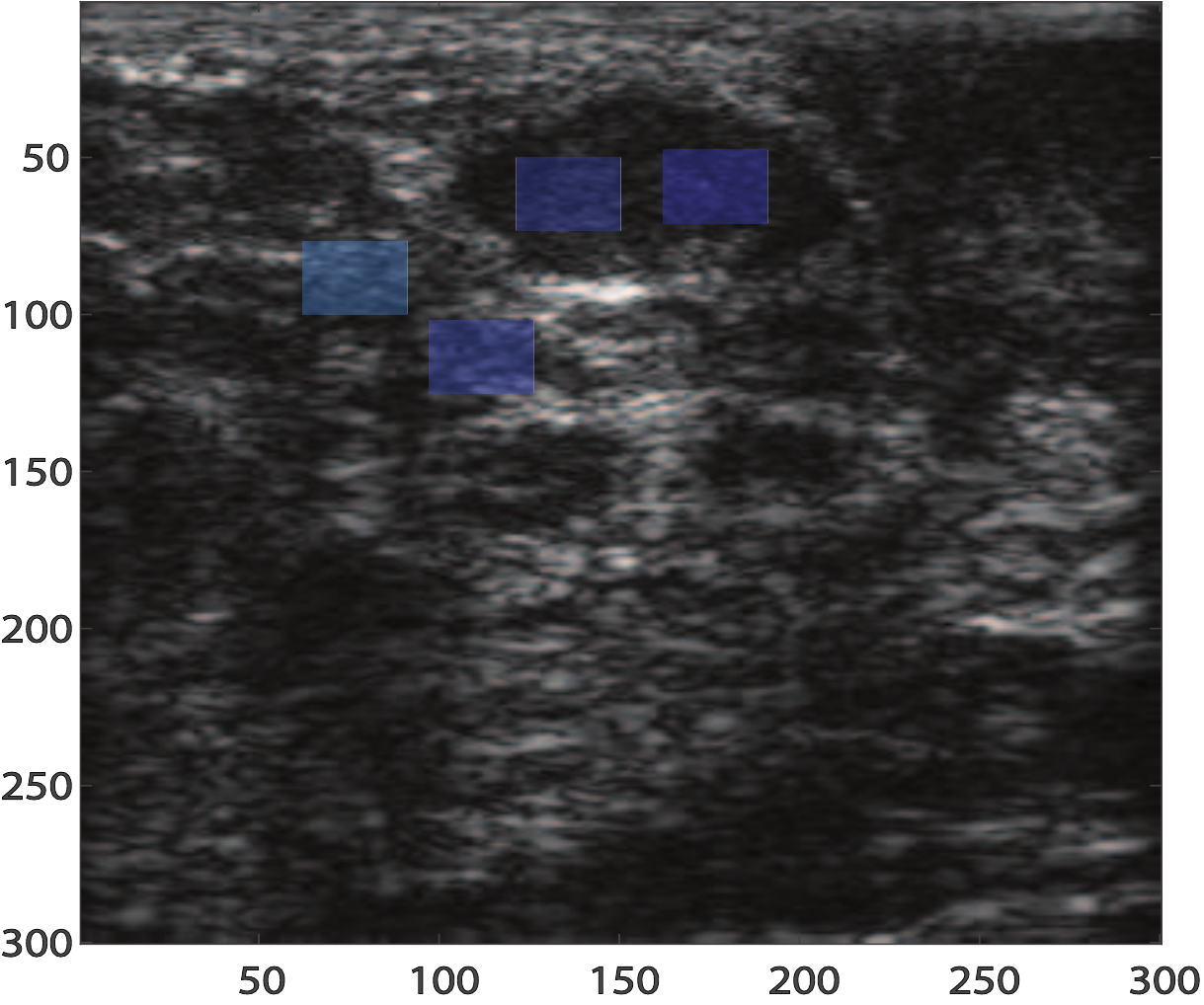}
\caption{ESD map for four ROIs for a representative fibroadenoma tissue. A deeper shade of blue indicates higher ESD values. The values on the axes are arbitrary.}
\end{figure}

The MSS estimation results for different types of breast tissues are presented in Table \ref{mss_res}.  The MSS values estimated for fibroadenomas, malignant lesions and normal breast tissues fall within the range of values previously reported  \cite{Bige_05}, \cite{Tadayyon_14}. As in the case of ESD, the MSS values of cysts and inflammatory lesions have not been reported before. The estimated MSS values for inflammatory lesions are slightly higher than those for normal tissues while the MSS values for cysts are inconsistent as expected and the algorithm fails to produce any reasonable estimates similar to the case for ESD estimation. Therefore, the average MSS value of cysts are not presented in the table. The average MSS value for ROIs taken outside the lesions is estimated to be $0.70$ ($\pm0.04$) mm, corresponding to that of normal tissues. The first and second rows of the table present the results for MSS estimation using a modified spectral autocorrelation (SAC) based technique \cite{Tadayyon_14} and a singular spectrum analysis (SSA) based technique \cite{Pereira_01}. These two techniques have been chosen, for comparison, because the modified SAC based technique \cite{Tadayyon_14} has been used for tissue characterization while the SSA based technique \cite{Pereira_01} is a signal decomposition based technique, not unlike our proposed MSS estimation technique. It is evident that, in both these techniques, the separation between the average MSS values of different groups of tissues is smaller with a higher SD.
\begin{table}[!ht]
\begin{center} \caption{Estimated values of MSS with SD (in bracket) using different methods for normal and pathological tissues.}
\label{mss_res}
\renewcommand{\arraystretch}{1.0}
\resizebox{\columnwidth}{!}{
\begin{tabular}{l|ccccc}
\hline {\footnotesize } & {\footnotesize  }
 & {\footnotesize  MSS ($\pm$SD) (mm)}& {\footnotesize  }\\
 {\footnotesize Method} & {\footnotesize  Malignant}
 & {\footnotesize  Fibroadenoma}& {\footnotesize  Normal} & {\footnotesize  Inflammatory}\\\hline
 {\footnotesize Modified SAC \cite{Tadayyon_14} } & \footnotesize   0.75 ($\pm$0.05)
 & \footnotesize   0.77 ($\pm$0.05) & \footnotesize  0.80 ($\pm$0.07)& \footnotesize  0.78 ($\pm$0.07)\\
 {\footnotesize SSA \cite{Pereira_01} }   & \footnotesize  0.87 ($\pm$0.05)
 & \footnotesize  0.83 ($\pm$0.05)   & \footnotesize  0.80 ($\pm$0.05)& \footnotesize  0.82 ($\pm$0.06)\\
  {\footnotesize Proposed Method   } & \footnotesize  0.79 ($\pm$0.04)
 & \footnotesize  0.75 ($\pm$0.03)& \footnotesize  0.69 ($\pm$0.03)& \footnotesize  0.73 ($\pm$0.04)\\
 \hline
\end{tabular}
}
\end{center}
\end{table}

\subsection{ Classification Results}

 Table \ref{ESD_class} presents the classification results based on only ESD for the proposed method, the MASD-based method \cite{Oleze_02}, and the frequency domain method \cite{Zachary_02}. For the proposed method, a classification scheme based on ESD alone produces sensitivity, specificity, and accuracy values of $91.07$\%, $96.12$\%, and $94.34$\%, respectively which is significantly better than the other techniques used for comparison in this paper \cite{Oleze_02}, \cite{Zachary_02}. Additionally, the $Sum_5$ and MCC values of $469.44$ and $0.8755$, respectively are clearly more superior compared to the other techniques. When the Fisher's exact test \cite{Fisher_92} is used for testing statistical significance, the proposed method achieved significantly better quality metrics than the MASD-based method \cite{Oleze_02} and the frequency-domain method \cite{Zachary_02} with $p$ value equal to $0.005$ and $0.00001$, respectively. The area under the receiver operating characteristic (ROC) curve calculated using a bootstraping strategy ($200$ bootstraps) to obtain the mean and $95$\% confidence intervals (CIs) of the ROC curve and AUC yields a mean AUC value of 0.94 with a CI of $0.90$-$0.98$ for classification based on our proposed ESD estimator.
 \begin{table*}[t]
\begin{center} \caption{ Breast lesion classification results using ESD for different methods.}
\label{ESD_class}
\renewcommand{\arraystretch}{2.0}
\resizebox{\textwidth}{!}{
\begin{tabular}{|c|c|c|c|c|c|c|c|c|c|c|c|}
\hline {\footnotesize \textbf{Method}} & {\footnotesize  \textbf{TP}   }  & {\footnotesize  \textbf{TN}   }& {\footnotesize  \textbf{FP}   }& {\footnotesize  \textbf{FN}   }& {\footnotesize  \textbf{Sens. }   }& {\footnotesize  \textbf{Spec.}   }& {\footnotesize  \textbf{Acc. } }& {\footnotesize  \textbf{PPV } } &{\footnotesize  \textbf{NPV } }&{\footnotesize  \textbf{$Sum_5$}  }&{\footnotesize  \textbf{MCC} }\\
 {\footnotesize \textbf{}} & {\footnotesize  \textbf{}   }  & {\footnotesize  \textbf{}   }& {\footnotesize  \textbf{}   }& {\footnotesize  \textbf{}   }& {\footnotesize  \textbf{(\%) }   }& {\footnotesize  \textbf{(\%) }   }& {\footnotesize  \textbf{(\%) } }& {\footnotesize  \textbf{(\%)} } &{\footnotesize  \textbf{(\%) } }&{\footnotesize  \textbf{}  }&{\footnotesize  \textbf{}}  \\ \hline
MASD \cite{Oleze_02} & 20 & 93 & 10 & 36 & 35.71 & 90.32 & 71.06 & 66.66 & 72.09 & 335.83 & 0.3175\\ \hline
Frequency Domain \cite{Zachary_02} & 29 & 85 & 18 & 27 & 51.79 & 82.52 & 71.69 & 61.70 & 75.89 & 343.60 & 0.3591\\ \hline
Proposed & 51 & 99 & 4 & 5 & 91.07 & 96.12 & 94.34 & 92.73 & 95.19 & 469.44 & 0.8755 \\ \hline
\end{tabular}
}
\end{center}

\end{table*}

 Table \ref{QUS_class} presents the classification performance of MSS for different methods. The first and second rows of the table presents the classification performance of the MSS values estimated from the modified SAC \cite{Tadayyon_14} and SSA \cite{Pereira_01} based techniques, respectively. The third row presents the results obtained using the modified technique used in this paper. We see that the classification results, obtained using are proposed technique, are superior compared to the other techniques \cite{Tadayyon_14}, \cite{Pereira_01}. However, the classification performance of MSS on this dataset is not as high as that of ESD. Next, the results of combining ESD with MSS is shown in Table \ref{combo}. It is observed from the table that when ESD is fused with MSS, we obtain improved sensitivity, specificity, accuracy, and MCC values of $96.43\%$, $95.15\%$, and $95.60\%$, and $0.9054$, respectively. The mean AUC value, in this case, is found to $0.96$ with a $95$\% CI of $0.91$-$0.98$. These performance metrics are clearly better than classification based on only a single QUS parameter (either the ESD or MSS). The first two rows of Table \ref{combo} again presents the classification results obtained using our proposed ESD and MSS estimation techniques, for reference.
 \begin{table*}[t]
\begin{center} \caption{ Breast lesion classification results obtained using MSS for different methods.}
\label{QUS_class}
\renewcommand{\arraystretch}{1.5}
\resizebox{\textwidth}{!}{
\begin{tabular}{|c|c|c|c|c|c|c|c|c|c|c|c|}
\hline {\footnotesize \textbf{Method}} & {\footnotesize  \textbf{TP}   }  & {\footnotesize  \textbf{TN}   }& {\footnotesize  \textbf{FP}   }& {\footnotesize  \textbf{FN}   }& {\footnotesize  \textbf{Sens. }   }& {\footnotesize  \textbf{Spec.}   }& {\footnotesize  \textbf{Acc.} }& {\footnotesize  \textbf{PPV} } &{\footnotesize  \textbf{NPV } }&{\footnotesize  \textbf{$Sum_5$}  }&{\footnotesize  \textbf{MCC}}\\
 & {\footnotesize  \textbf{}   }  & {\footnotesize  \textbf{}   }& {\footnotesize  \textbf{}   }& {\footnotesize  \textbf{}   }& {\footnotesize  \textbf{(\%) }   }& {\footnotesize  \textbf{(\%) }   }& {\footnotesize  \textbf{(\%) } }& {\footnotesize  \textbf{(\%)} } &{\footnotesize  \textbf{(\%) } }&{\footnotesize  \textbf{}  }&{\footnotesize  \textbf{}}\\\hline
Modified SAC \cite{Tadayyon_14} & 37 & 55 & 48 & 19 & 66.07 & 53.40 & 57.86 & 43.53 & 74.32 & 295.18 & 0.1864\\ \hline
SSA \cite{Pereira_01} & 20 & 89 & 14 & 36 & 35.71 & 86.41 & 68.55 & 58.82 & 71.20 & 320.70 & 0.2577\\ \hline
Proposed & 42 & 72 & 31 & 14 & 75 & 69.90 & 71.70 & 57.53 & 83.72 & 357.86 & 0.4304 \\
								
 \hline
\end{tabular}
}
\end{center}

\end{table*}
 As discussed before, the classification performance of the combination of ESD and MSS has been evaluated directly using SVM, LDA, MNR, KNN, and Na\"ive Bayes classifiers. In Table \ref{combo}, the reported $Sum_5$ value of $476.69$ is obtained for a quadratic LDA classifier and represents the best $Sum_5$ value obtained out of all the above mentioned classifiers. The average $Sum_5$ value ($\pm$SD), found by averaging the $Sum_5$ values obtained from each of the mentioned classifiers, is $473.02$ ($\pm$ $3.47$). This indicates that the classification performance obtained using the combination of ESD and MSS is fairly stable.

 \begin{table*}[t]
\begin{center} \caption{ Breast lesion classification results obtained using a combination of ESD and MSS.}
\label{combo}
\renewcommand{\arraystretch}{1.5}
\resizebox{\textwidth}{!}{
\begin{tabular}{|c|c|c|c|c|c|c|c|c|c|c|c|}
\hline {\footnotesize \textbf{Parameter}} & {\footnotesize  \textbf{TP}   }  & {\footnotesize  \textbf{TN}   }& {\footnotesize  \textbf{FP}   }& {\footnotesize  \textbf{FN}   }& {\footnotesize  \textbf{Sens. }   }& {\footnotesize  \textbf{Spec.}   }& {\footnotesize  \textbf{Acc.} }& {\footnotesize  \textbf{PPV} } &{\footnotesize  \textbf{NPV } }&{\footnotesize  \textbf{$Sum_5$}  }&{\footnotesize  \textbf{MCC}}\\
 & {\footnotesize  \textbf{}   }  & {\footnotesize  \textbf{}   }& {\footnotesize  \textbf{}   }& {\footnotesize  \textbf{}   }& {\footnotesize  \textbf{(\%) }   }& {\footnotesize  \textbf{(\%) }   }& {\footnotesize  \textbf{(\%) } }& {\footnotesize  \textbf{(\%)} } &{\footnotesize  \textbf{(\%) } }&{\footnotesize  \textbf{}  }&{\footnotesize  \textbf{}}\\\hline
ESD  & 51 & 99 & 4 & 5 & 91.07 & 96.12 & 94.34 & 92.73 & 95.19 & 469.44 & 0.8755\\ \hline
MSS & 42 & 72 & 31 & 14 & 75 & 69.90 & 71.70 & 57.53 & 83.72 & 357.86 & 0.4304\\ \hline
ESD and MSS & 54 & 98 & 5 & 2 &  96.43 & 95.15 & 95.60 & 91.53 & 98.00 & 476.69 & 0.9054\\ \hline

 \hline
\end{tabular}
}
\end{center}

\end{table*}
Finally, we investigate how the different steps of the proposed algorithm impact the classification accuracy when only ESD is used  for classifying the breast lesions. The results are shown in Table \ref{classification_ablation}. It is clear from the table that removing the EEMD step has the most drastic impact on classification performance with a sharp fall in the sensitivity, specificity, and accuracy results from the proposed method. Furthermore, removal of the exponential neighborhood step has a significantly detrimental impact on the classification performance. This can be correlated to the earlier observations, from experimental TMPs and \textit{in vivo} tissues, that removing the weighted neighborhood step significantly increases the SD of the ESD estimates. It is also clear that removing any one of the system effect minimization steps also impacts the classification performance adversely. This again confirms the idea that system effect minimization requires a multi-step approach. Therefore, the different steps of the proposed ESD estimation algorithm indeed contributes to improving the classification performance.  Furthermore, the use of $Sum_5$ to indicate the classification performance has been previously established in \cite{Nizam_17}, \cite{Ara_2017}, \cite{Ara_15}. In this paper, we have also used MCC to evaluate the classification performance which is shown to have a high degree of correlation with the $Sum_5$ values. That is, a higher $Sum_5$ value generally leads to an MCC value closer to +$1$ with a minor exception, which is observed, if we compare the `No Normalization' and `No Deconvolution' cases in Table \ref{classification_ablation}.
\begin{table*}[!ht]
\begin{center} \caption{ Breast lesion classification results obtained using only ESD by removing the different steps in the proposed technique.}
\label{classification_ablation}
\renewcommand{\arraystretch}{2.0}
\resizebox{\textwidth}{!}{
\begin{tabular}{|c|c|c|c|c|c|c|c|c|c|c|c|}
\hline {\footnotesize \textbf{Condition}} & {\footnotesize  \textbf{TP}   }  & {\footnotesize  \textbf{TN}   }& {\footnotesize  \textbf{FP}   }& {\footnotesize  \textbf{FN}   }& {\footnotesize  \textbf{Sens. }   }& {\footnotesize  \textbf{Spec.  }   }& {\footnotesize  \textbf{Acc. } }& {\footnotesize  \textbf{PPV} } &{\footnotesize  \textbf{NPV } }&{\footnotesize  \textbf{$Sum_5$}  }&{\footnotesize  \textbf{MCC}}\\
{\footnotesize \textbf{}} & {\footnotesize  \textbf{}   }  & {\footnotesize  \textbf{}   }& {\footnotesize  \textbf{}   }& {\footnotesize  \textbf{}   }& {\footnotesize  \textbf{(\%) }   }& {\footnotesize  \textbf{(\%) }   }& {\footnotesize  \textbf{(\%) } }& {\footnotesize  \textbf{(\%)} } &{\footnotesize  \textbf{(\%) } }&{\footnotesize  \textbf{}  }&{\footnotesize  \textbf{}}\\ \hline
Proposed  & 51 & 99 & 4 & 5 & 91.07 & 96.12 & 94.34 & 92.73 & 95.19 & 469.44 & 0.8755\\ \hline
No Filtering & 49 & 97 & 6 & 7 & 87.50 & 94.17 & 91.82 & 89.09 & 93.27 & 455.86 & 0.8202\\ \hline
No Normalization & 51 & 54 & 9  & 5 & 91.07 & 91.26 & 91.19 & 85.00 & 94.95  & 453.48& 0.7666 \\ \hline
No Deconvolution & 49 & 96 & 7 & 7 & 87.50 & 93.20 & 91.19 & 87.50 & 93.20 & 452.60 & 0.8070\\ \hline
No Neighborhood & 39  & 100 & 3 & 17 & 69.64 & 97.08 & 87.42 & 92.86 & 85.47 & 432.48& 0.7230 \\ \hline
No EEMD & 43 & 81 & 22 & 13  & 76.78 & 78.64 & 77.99  & 66.15 & 86.17 & 385.74 & 0.5385\\ \hline

 \hline
\end{tabular}
}
\end{center}
\end{table*}

\section{Discussion}
In this paper, we proposed a mutli-QUS parameter based breast lesion classification scheme using ESD and MSS. In order to obtain reliable ESD estimates for breast lesion classification, an improved ESD estimation technique for breast tissues has been proposed based on the separation of the diffuse component from the coherent component using EEMD and a multi-step system effect minimization technique. Also, a nearest neighborhood algorithm has been applied to fit an average regression line in the frequency domain, from which the ESD is estimated. The proposed ESD estimator has been shown to outperform the conventional techniques, which employ a Gaussian form factor as exhibited by a lower MAPE value in Table \ref{TMP}. An ablation technique has been applied to show the suitability of a multi-step system effect minimization technique in Fig. 5 and Table \ref{classification_ablation} . The use of a weighed exponential neighborhood for an average regression line fitting has also been justified through ablation in Table \ref{classification_ablation} and and an ESD map in Fig. 6. The higher values of the performance metrics in Table \ref{ESD_class} compared to the other techniques show the suitablity of the proposed ESD estimator to be used as a breast lesion classification tool. Table \ref{combo} demonstrates the suitability of the proposed ESD and MSS estimators to be used as CAD tools for breast lesion classification, when they are used in conjunction. This opens up an avenue for the use of QUS parameters like the ESD and MSS as non-invasive quantitative biomarkers for breast cancer detection. However, it is to be noted that values of ESD and MSS can not be estimated for cystic lesions. But, as cysts are classified as benign lesions, a  ESD/MSS based classification scheme will be unable to characterize cysts correctly. Hence, this is clearly a limitation of such a classification scheme.  However, some strain imaging techniques have been developed which can successfully characterize cysts \cite{Varghese_96}, \cite{Rabbi_17} and hence, may be used in conjunction with ESD and MSS for breast lesion classification. Furthermore, it is clear from (\ref{3_def_c}) that the proposed technique is able to estimate another QUS parameter, EAC. But, the EAC values estimated using this technique have a poor classification performance on this dataset. Thus, a careful revision of the proposed method may be required to get insights into the poor performance of an EAC based classifier. In addition, this paper employs a Gaussian form factor to model tissue scattering. Although, this has produced reasonable ESD estimates in this work, similar schemes using other form factors such as the Faran form factor, need to be implemented before an informed decision can be made about which form factor best models scattering from breast tissues for QUS parameter estimation. Also, the true gold-standard for QUS parameters relating to tissue micro-structures like the ESD and MSS has to be obtained using microscopy analysis from histopathology slides. Because of a lack of histopathology slides for our dataset, we could not carry out such an analysis. If future works can correlate ESD  to breast tissue histopathology, it may be more reliably employed as a non-invasive tissue marker for breast cancer detection. Moreover, recent trends suggest the use of deep neural networks for ultrasonic tissue characterization \cite{Han_17}. The dataset used in this work is still too limited for such an approach. Future works, on a larger dataset, may look to combine the feature based approach of conventional QUS classifiers with the data-driven approach employed in deep neural networks for a more robust classification scheme.

\section{Conclusion}
This paper has presented a multi-QUS biomarker based breast lesion classification scheme which uses ESD and MSS as the QUS parametes. An improved frequency domain technique for ESD estimation from the diffuse component of the backscattered data using EEMD has also been proposed. The ESD estimates based on this technique has been shown to produce a better classification perfomance compared to some of the existing techniques for ESD estimation. Additionally, when the ESD is fused with another prominent QUS parameter, the MSS, estimated from an ameliorated EEMD domain AR spectral estimation technique, the classification performance further improves. Therefore, the proposed ESD estimator, in conjunction with the proposed MSS estimator, show promises to be used as CAD tools for breast lesion classification. Also, it opens the possibility for the use of these QUS parameters as non-invasive biomarkers for breast cancer detection.

\section{Acknowledgement}
This work was supported by HEQEP UGC, Bangladesh, under Grant CPSF\#96/BUET/Win-2/ST(EEE)/2017. The \textit{in vivo} breast data were acquired at BUET Medical Center by Dr. F. Alam, Assistant Professor of Radiology and Imaging, Bangabandhu Sheikh Mujib Medical University, Dhaka-1000, Bangladesh. The computer programs for the methods in \cite{Oleze_02} and \cite{Zachary_02} were derived from the graphical user interface (GUI) developed by the authors of those papers. We would like to thank the authors for providing us with the GUI.
\bibliographystyle{IEEEtran}
\bibliography{bibliography_ESD}

\end{document}